\documentclass[aps,pra,showpacs,twocolumn]{revtex4-1}
\usepackage{amssymb}
\usepackage{amsfonts}
\usepackage{amssymb}
\usepackage{amsmath}
\usepackage{graphicx}
\usepackage{subfigure}

\makeatletter

\newcommand{\Rmnum}[1]{\expandafter\@slowromancap\romannumeral #1@}
\makeatother

\begin{document}

\title{Interaction and filling-induced quantum anomalous Hall effect in ultra-cold neutral Bose-Fermi mixture on hexagonal lattice}
\author{Shang-Shun Zhang, Heng Fan, Wu-Ming Liu}
\affiliation{ Beijing National Laboratory for Condensed Matter Physics, Institute of
Physics, Chinese Academy of Sciences, Beijing 100190, China }

\begin{abstract}

We investigate the quantum anomalous Hall effect in a mixture of ultra-cold neutral bosons and fermions held on a hexagonal optical lattice. In the strong atom-atom interaction limit, composite fermions composed of one fermion with bosons or bosonic holes in the mixture are formed. Such composite fermions have already been generated successfully in experiment [Nat. Phys. {\bf 7}, 642 (2011)]. Here we predict that this kind of composite fermions may provide a realization of the quantum anomalous Hall effect by tuning the atom-atom interaction or the filling of the bosons in the mixture. We also discuss the corresponding experimental signatures of the quantum anomalous Hall effect in the Bose-Fermi mixture on hexagonal optical lattice.

\end{abstract}

\maketitle

\section{Introduction}
Recently, quantum anomalous Hall (QAH) effect and closely related topics have been attracting a great deal of attentions since of their fundamental interests and potential technological applications, such as new generation of quantum electronic devices \cite{youruiscience,Motohiko,HuaJiang,Hongbin,Hongbin2,JanMarl}. In QAH effect, the quantized value of Hall conductance is related to a bulk topological number and is robust against disorder and other perturbations. This non-trivial topology is guaranteed by the breaking of the time reversal symmetry. In contrast to the quantum Hall effect which happens at strong magnetic field and low enough temperature, the QAH effect is induced without any external magnetic field applied to the system \cite{Halleffect}. In the past years, the QAH effect in solid state systems has been studied both theoretically and experimentally. It is theoretically predicted that this intrinsic quantum Hall effect is realizable in semiconductor systems \cite{youruiscience,cxliuquantumwell,Ferromagnetic} and in graphene  \cite{qiao2011,qiaoprb,qiao2010,Taillefumier}. However, it has not been observed in experiment so far.

On the other hand, cold atom physics has been extended to many domains such as statistical physics, condensed matter, and quantum information and provides a tunable artificial platform to study various novel quantum phenomena \cite{rmpcoldatom,Bloch,StevenChu,Jaksch}. The experimental schemes to realize the quantum Hall effect and the quantum anomalous Hall effect by cold atoms have been proposed. Those methods include globally rotating the trap or optical lattice, or introducing synthetic gauge potential generated by laser beams and so on \cite{Chuanwei,Goldman,Kun,Baranov,Susanne}. The QAH effect is also predicted to be realizable in the $p$-band optical lattice system by rotating each optical lattice site around its own center \cite{wucjprl,Braggcjwu}. Yet, in contrast to the experimental discovery of the quantum Hall \cite{QHKlitzing} and quantum spin Hall effects \cite{SCScience,MKScience}, no observation of QAH effect in condensed-matter or cold-atom systems has been reported.

In this paper, we present a practical scheme to realize the QAH in the Bose-Fermi mixture on hexagonal lattice. It is pointed out that in the strong interaction limit, different composite fermions can be formed in the Bose-Fermi mixture on optical lattice \cite{Lewenstein}. In this process, the free fermion combines with bosons to form the composite fermion when they are strongly coupled. The composite fermions can interact with each other on neighbor sites, and the strength of the interaction can be controlled by the filling of bosons. In a recent experiment, different kind of composite fermions and phase separation have already been observed \cite{Sugawa}. We know that the system of hexagonal lattice provides a platform to study Dirac fermion, which leads to various novel phenomena under different conditions, including  such as the topological none-trivial state \cite{KMmodel,Panahi,Szirmai,Bermudez}. Motivated by these experimental developments and the rich properties of Bose-Fermi mixture, it is naturally to wonder whether such as QAH effect can be realized by Bose-Fermi mixture on hexagonal lattice. This scheme is indeed possible and is compatible with current experiment techniques. When the Dirac fermion on the hexagonal lattice is strongly coupled with the bosons, the nearest neighbor (NN) and next nearest neighbor (NNN) interactions between composite fermions are generated and controlled by the filling of bosons. The competition of the NN and the NNN interactions drives the Bose-Fermi mixture to either the QAH regime or charge density wave, as pointed out in Ref. \cite{Raghu}. Differing from the previous proposed schemes based on the spin-orbit coupling, our proposal offers an alternative method to realize the QAH effect in the rapidly developing system of Bose-Fermi mixture  \cite{Titvinidze,Dutta,Buchler,Bruderer,Sugawa} and provides a real system to test our understanding of the essence of the QAH effect.

The paper is organized as follows. We first describe the model building of the Bose-Fermi mixture held on hexagonal lattice in Sec. \Rmnum{2}. The concept of the composite fermion and the interaction and filling effect on the composite fermion is presented in the strong interaction limit. In Sec. \Rmnum{3}, the effective Hamiltonian of the composite fermion on hexagonal lattice is discussed. Based on the analysis of the mixture, we turn to discuss the realization of QAH effect in Sec. \Rmnum{4}. The experimental signatures are discussed in Sec. \Rmnum{5}.
\section{Bose-Fermi mixture on hexagonal lattice}

\begin{figure}[t]
\begin{center}
\includegraphics[width=3in]{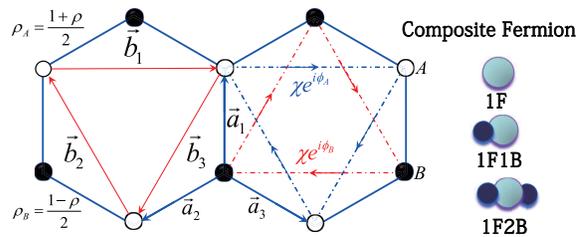}\hspace{0.5cm}
\caption{
(Color online) Left panel: Schematic of the hexagonal lattice geometry structure and
the mean-filed channels for composite fermion.
The lattice contains two triangle sublattices, labeled by $A$ (open circle) and
$B$ (black ball). The vector $\vec a_i,i=1,2,3$ are the nearest-neighbor displacements
from site $B$ to site $A$. And vector $\vec b_i,i=1,2,3$ are second-neighbor displacements.
$\rho_A$ and $\rho_B$ denotes the particle density at site $A$ and $B$. $\chi e^{i \phi_{A,B}}$
is the second neighbor sites hopping driven by the second neighbor interaction.
Right panel: Schematic of the composite fermions for three types: one fermion, one fermion with one boson, one fermion with two bosons.
}
\label{lattice}
\end{center}
\end{figure}

We consider the Bose-Fermi mixture on the hexagonal lattice. The Hamiltonian is written as
\begin{align}
H_{BFH}& =-\sum \limits_{\langle ij\rangle }(
J_{B,1}b_{i}^{\dag}b_{j}+J_{F,1}f_{i}^{\dag}f_{j}+h.c.) \notag  \\
& -\sum \limits_{\langle \langle ij\rangle \rangle }(
J_{B,2}b_{i}^{\dag}b_{j}+J_{F,2}f_{i}^{\dag}f_{j}+h.c.) \notag  \\
& +\sum \limits_{i}[ \frac{1}{2}Vn_{i}( n_{i}-1) -\mu n_{i}%
] +U\sum \limits_{i}n_{i}m_{i},
\end{align}
where $b_{i}$ and $f_{i}$ represent boson and fermion annihilation operators on site $i$, and $n_{i}=b_{i}^{\dag}b_{i}$ and $ m_{i}=f_{i}^{\dag}f_{i}$ are the corresponding particle number operators. The first term represents the nearest neighbor hopping on the hexagonal lattice with $J_{F,1}$ and $J_{B,1}$ as the corresponding hopping parameters for fermion and boson respectively. The second term represents the next nearest neighbor hopping on the hexagonal lattice, and similarly $J_{F,2}$, $J_{B,2}$ are the corresponding hopping parameters.  Fig. \ref{lattice} gives a schematic of the hexagonal lattice, which contains two sets of sublattices illustrated in the caption. The on-site interaction between bosons is described by $V$, which is assumed to be repulsive. The inter-species interaction between bosons and fermions is $U$ , which can be either positive or negative. There is no on-site interaction between spinless fermions due to the Pauli exclusion principle. $\mu $ is the chemical potential for bosons. The chemical potential for fermions is absent because the filling of fermions is assumed to be fixed, which equals $1/2$.

In the following, we consider the strong interaction limit, $U,V \gg J$. Ref. \cite{Lewenstein} has investigated the case without assuming the geometry and dimensionality of the lattice. Let us first recall the main results in Ref. \cite{Lewenstein} briefly. In the limit of strong on-site intra-species boson-boson interaction $V$, the particle number of the bosons per site in the ground state is determined by $n=\left[ \widetilde{\mu} \right] +1$, where $\widetilde{\mu}$ is the dimensionless chemical potential defined as $\widetilde{\mu}=\mu/V $ and $\left[ \widetilde{\mu} \right] $ labels the integer part of $\widetilde{\mu}$. The strong on-site inter-species boson-fermion interaction $U$ leads to the formation of boson-fermion composite particle. In the strong interaction limit, the composite particle can be considered as an entity. It is described by the composite fermion annihilation operator
\begin{equation}
\tilde{f}_{i}=\sqrt{(n-s) !/n!}( b_{i}^{\dag}) ^{s}f_{i}, s>0,
\end{equation}
or
\begin{equation}
\tilde{f}_{i}=\sqrt{(n-s) !/n!}( b_{i}) ^{-s}f_{i}, s<0,
\end{equation}
where $s<0$ ($s>0$) represents number of bosons (holes) in the composite fermion. $s$ is determined by $\widetilde{\mu}$ and $\alpha$, where $\alpha$ is the dimensionless interaction ratio $U/V$. It should be noted that  $s\le n$, which is due to the fact that the maximum particle number of the bosons per site is $n$ in the ground state.

\begin{figure}[t]
\centering
\includegraphics[width=3in]{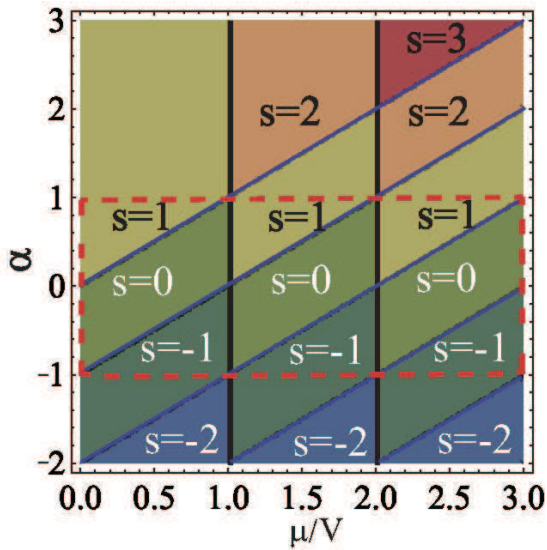}\hspace{0.5cm}
\caption{(Color online) The phase diagram for ground state of the Bose-Fermi mixture in the limit $J \rightarrow 0$ on the $\widetilde{\mu}$-$\alpha$ plane. The blocks with different colors represent the regions with different composite particles formed, which is determined by Eq. \ref{eq:Ns}. For each block, the ground state of the system is characterized by the type of the composite particle with $s$ and the number of the occupation of bosons $n$. The region marked by red dashed line is the region we considered in this work.}
\label{fig:block}
\end{figure}

There are two key parameters relevant to the physical results of our model and both of them could be tuned in experiment. One is the dimensionless chemical potential $\widetilde{\mu}$ which is defined above. And the other one is the interaction ratio $\alpha=U/V$. The structure of the composite fermion is determined by
\begin{equation}\label{eq:Ns}
s=\left[  \alpha-\widetilde{\mu}+\left[  \widetilde{\mu} \right]  \right]  +1.
\end{equation}
So different composite fermions can be generated by tuning the controllable parameters $\widetilde{\mu}$ and $\alpha$. To give a visual description of composite particle, we show the structure of the composite fermions on the $\widetilde{\mu}$-$\alpha$ plane in Fig. \ref{fig:block}. In this graph, we find that the $\widetilde{\mu}$-$\alpha$ plane is divided into several blocks with different composite particle formed. On each of the block, the structure of the composite particles is independent to the interaction ratio $\alpha$ and dimensionless chemical potential $\widetilde{\mu}$. For the region with $\alpha>\widetilde{\mu}$, all of the bosons are pushed out of the lattice site by fermions, i.e., there is $s=n$. Fig. \ref{fig:block} gives the phase diagram for ground state of the Bose-Fermi mixture for deep optical lattice, namely $J_{F},J_{B} \rightarrow 0$. The phase diagram in the weak hopping limit is independent to the geometry and the dimensionality of the optical lattice. However, more interesting phenomena appear when the hoppings of the atoms are turned on. After turning on the hoppings of the particles, the concept of the composite particle still makes scenes if the strength of hoppings is much weaker than the inter-particles interaction. For this case, the phase of the system can also be naturally characterized by these blocks in Fig. \ref{fig:block}. In addition, new boundaries for different phases of the system appear in each of the blocks, and the details depend on the geometry and dimensionality of the optical lattice, which will be discussed in the following sections.

\section{The effective Hamiltonian for composite fermions on hexagonal lattice}
In this section, we consider the effect of the hopping terms. In the limit $J \rightarrow 0$, the ground state of the system is Mott insulator, with each of the lattice site contains $\widetilde{n}$ bosons and at most one composite fermion. For an optical lattice with $N$ sites and $m$ spinless fermions loaded (the filling of bosons is determined by $\mu$), the degeneracy of the ground state is $N!/m!(N-m)!$. In the case with the filling fraction of fermions equals $1/2$, the degeneracy has the asymptotic behavior as $2^{N+1}/\sqrt{N}$ for $N \rightarrow \infty$. The energy gaps for the excited levels have the order of $U,V$. Therefore the distribution on the excited states can be neglected for low temperature $k_BT \ll U,V$. The hopping terms provide a periodic perturbation for the system, and the degeneration will be lifted, and therefore the degenerated levels will form an energy band. The bandwidth will be much narrower than the energy gaps of these excited levels for weak hopping terms $J \ll U,V$. Thus we can limit our consideration to the lowest energy band in this case.

With the method of the derivation of t-J model, we can obtain the effective Hamiltonian limited in the lowest energy band. The hopping terms are assumed to be weak compared to the interaction strength, such that we could expand the effective Hamiltonian relative to the hopping parameters order by order. As analyzed above, the composite particles behave as entities, which gives rise to conclusion that the lowest order for the effective interactions between neighbor sites is $2$ and the effective hopping terms is $|s|+1$ \cite{Lewenstein}. Therefore we can estimate that the strength of the effective interactions are  about $J^2/U$ and the effective hopping parameters are about $J^{|s|+1}/U^{|s|}$. After tuning on the hopping terms, the degenerated energy levels of the ground state of the system form an energy band, the bandwidth of which has the same order with the effective hopping parameters. Since the effective hopping parameters estimated as $J^{|s|+1}/U^{|s|}$ is very weak for $|s| \gg 1$, the energy band is extremely flat and similar to case for $J \rightarrow 0$. The zero temperature approximation for the behaviors of the atoms lying on the lowest energy band is not proper. And the predictions based on the ground state properties of the lowest energy band will break down. Therefore we choose the region with $s=0$ in the following. The possible region has been marked by red dashed line in Fig. \ref{fig:block}.

The effective Hamiltonian for neighbor interaction can be obtained by the second order perturbation:
\begin{equation}
V_{eff}^{( 2) }=U_{eff}\sum \limits_{\langle ij\rangle}\tilde{m}%
_{i}\tilde{m}_{j}+V_{eff}\sum \limits_{\langle \langle ij\rangle \rangle}%
\tilde{m}_{i}\tilde{m}_{j}, \label{eff_interaction}
\end{equation}
where $\tilde{m}_i=\tilde{f}_i^{\dagger} \tilde{f}_i$. For $s=0$, the effective parameters are
\begin{subequations}\label{UV}
\begin{eqnarray}
U_{eff} & =\frac{J_{B1}^{2}}{V} \frac{4 \alpha^2}{1-\alpha^2} n (n+1),\\
V_{eff} & =\frac{J_{B2}^{2}}{V} \frac{4 \alpha^2}{1-\alpha^2} n (n+1).
\end{eqnarray}
\end{subequations}
The effective interactions between the composite particles are induced by the hopping of the bosons. As a result of second order perturbative expansion, the effective interactions have the leading order $J^2/V$, however, they are tunable via the interaction ratio $\alpha$ and filling of bosons $\widetilde{\mu}$. For each block with fixed $s$ and $n$, the effective parameters are independent to the $\mu$, while they are related to the interaction ratio through $4 \alpha^2/(1-\alpha^2)$. The occupation of bosons $n$ on each site is determined by $\widetilde{\mu}$ via $n=[\widetilde{\mu}]+1$. Hence in the process of increasing the filling of bosons with $\alpha$ fixed, the curves of $U_{eff}$ and $V_{eff}$ have step-like forms.

The effective hopping term is given by the $|s|+1$-th order perturbation:
\begin{equation}
T_{eff}^{( |s|+1) }=-J_{eff}\sum \limits_{\langle ij\rangle}\tilde{%
f}_{i}^{\dag}\tilde{f}_{j}-I_{eff}\sum \limits_{\langle \langle
ij\rangle \rangle}\tilde{f}_{i}^{\dag}\tilde{f}_{j} \label{eff_hopping}.
\end{equation}
For the case $s=0$ considered in this paper, the parameters read as $J_{eff} =J_{F,1}$, $I_{eff} =J_{F,2}$, which are determined by the hopping parameters of the fermions. We can see that the result is also independent to the filling of bosons $\widetilde{\mu}$. Therefore we conclude that the phase of the system in each block with fixed $s$ and $n$ are independent to $\widetilde{\mu}$. The phase boundary should be lines parallel to the $\widetilde{\mu}$ axes as shown in Fig. \ref{phase}. This character of the phase diagram is valid for weak enough hoppings of bosons and fermions compared to the atom-atom interaction. In practice, the hopping of fermions and bosons are finite values, and the phase boundaries might deviate from the parallel lines slightly.

\section{The quantum anomalous Hall effect in Bose-Fermi mixture
on hexagonal lattice}
The effective model given in Eq. (\ref{eff_hopping}) includes a NNN hopping term $-I_{eff}\sum
\tilde{f}_{i}^{\dag}\tilde{f}_{j}$ determined by hopping parameters of fermion.
It can be suppressed for weak second neighbor hopping $J_{F2}$.
Within this constraint, the effective Bose-Fermi model reduces to:
\begin{align}
H_{eff}  &  =-J_{eff}%
{\textstyle \sum \limits_{\langle ij\rangle}}
\tilde{f}_{i}^{\dag}\tilde{f}_{j}+U_{eff}%
{\textstyle \sum \limits_{\langle ij\rangle}}
\tilde{m}_{i}\tilde{m}_{j}+V_{eff}%
{\textstyle \sum \limits_{\langle \langle ij\rangle \rangle}}
\tilde{m}_{i}\tilde{m}_{j}, \label{effective_H}
\end{align}
where $\tilde{m}_i=\tilde{f}_i^{\dagger} \tilde{f}_i$. Due to the competition of the NN interaction and the NNN interaction in this effective Hamiltonian, the system shows up the CDW and orbital ordering phase \cite{Raghu,Honerkamp1,Honerkamp2}. The mean field configuration of CDW phase is described by a density fluctuation $\rho=\frac{1}{2}(  \langle f_{iA}^{\dag}f_{iA}\rangle-\langle f_{iB}^{\dag}f_{iB}\rangle ) $, which is shown in Fig. \ref{lattice}. The subscript $A$ and $B$ denote the two triangle sublattices of the hexagonal lattice described in the caption of Fig. \ref{lattice}. The orbital ordering phase is described by a complex second neighbor hopping parameter: $\langle f_{i}^{\dag}f_{j}\rangle=\chi_{ij}=\chi_{ji}^{\ast}$ (Fig. \ref{lattice}). Where $i,j$ is the second neighbor site. Since the system has the translational symmetry and rotational symmetry $C_3$ which are invariant for different quantum phases, the ansatz of mean field $\chi_{ij}$ is chosen as
$$
\chi_{i,i+\vec{b}}=
\left \{\begin{array}
[c]{c}%
\chi_{A}=\left \vert \chi \right \vert e^{i\phi_{A}},i\in A;\\
\chi_{B}=\left \vert \chi \right \vert e^{i\phi_{B}},i\in B.
\end{array}
\right.
$$
where $\vec{b}$ is a vector defined on the triangle sublattice. The mean field Hamiltonian reads as,
\begin{align} \label{eq:meanfield}
H_{eff}&=-J_{eff}%
{\textstyle \sum \limits_{\langle ij\rangle}}
\tilde{f}_{i}^{\dag}\tilde{f}_{j}+\sum \limits_{\langle \langle ij\rangle \rangle }
[V_{eff}\left \vert \chi \right \vert e^{i\phi_{i}}\tilde{f}_{i}^{\dag}\tilde{f}_{j}+h.c.]\notag \\
& + \epsilon_{A}
{\textstyle \sum \limits_{i\in A}}
\tilde{m}_{i}+\epsilon_{B}
{\textstyle \sum \limits_{i\in B}}
\tilde{m}_{i},
\end{align}
where $\epsilon_{A}=\rho U_{eff}$, $\epsilon_{B}=-\rho U_{eff}$. The exponential factor $e^{i \phi_i}$ in the second term corresponds to the orbital ordering phase, which induces the local
magnetic flux and breaks the time reversal symmetry spontaneously. The Hamiltonian of the mixture for
\begin{figure}[t]
\centering
\includegraphics[width=3.2in]{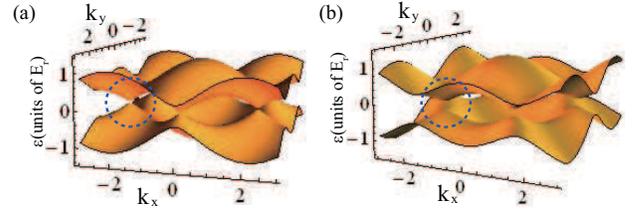}\hspace{0.5cm}
\caption{
(Color online) (a) The spectrum corresponding to the semimetal state. The spectrum keeps gapless at the Dirac points as marked by the the red dashed circle. (b) The spectrum corresponding to the QAH state. The spectrum at the Dirac points opens a gap when the filling of the bosons exceeding the critical value (marked by the blue dashed circle). The dimensionless chemical potential for bosons $\widetilde{\mu}$ is taken as $4.8$ and the other parameters are the same with Fig. \ref{fig:mass1}.
}
\label{spectrum}
\end{figure}
this case is mapped into the Haldane model \cite{Haldanemodel}. The last two terms break the symmetry of the two sublattices $A$ and $B$ due to the background of density wave fluctuation. The fermion has been assumed to be half filled. For $\rho=0$ and $\chi=0$, the spectrum of the system is shown by Fig. \ref{spectrum} (a), which owns two Dirac cones. The effective interaction induced by the hopping of bosons might drive the system to the ground state with non-zero values of $\rho$ and $\chi$. For this case, the spectrum of the system will open a gap at the two Dirac points as shown in Fig. \ref{spectrum} (b). The topology of the ground state is unchanged as long as the energy gap is kept open. Therefore, the mass terms at the two Dirac points are essentially important for the topological properties which are related to the appearance of QAH effect. Therefore, we try to obtain the linear expansion of the mean field Hamiltonian in Eq. (\ref{eq:meanfield}) around the two Dirac points, which read as: $H_{K/K^{\prime}}=-\frac{3 J_{eff}}{2} k_x\sigma_x-\frac{3 J_{eff}}{2}k_y\sigma_y+m^{\pm}\sigma_z$, where
\begin{align} \label{eq:mass}
m^{\pm}   &  =U_{eff}\rho \pm \frac{3\sqrt{3}}{2}V_{eff}\left \vert
\chi \right \vert \left( \sin \phi _{A}-\sin \phi _{B}\right)   \notag \\
& \quad -\frac{3}{2}V_{eff}\left \vert \chi \right \vert \left( \cos \phi
_{A}-\cos \phi _{B}\right).
\end{align}
Here $+$, $-$ correspond to the two Dirac points $K$ and $K^{\prime}$ respectively.
For $\rho=0$ and $\chi \neq 0$ (there is always $\phi_A$=-$\phi_B$=$\pi/2$), the signs for the
mass terms at the two Dirac points are different. However for $\chi=0$ and $\rho \neq 0$, the two
mass terms have the same sign. This sign difference for the two cases is topologically distinctive as long as the energy gaps at the two Dirac points are kept open, and is
captured by the first Chern number, which reads as
\begin{align} \label{Chern}
C_{1}=\frac{1}{2}\left(  \operatorname{sign}\left(  m^{+}
\right)  -\operatorname{sign}\left(  m^{-}  \right)  \right).
\end{align}
The first Chern number defined for the ground state of the insulator represents the Berry phase
of the adiabatic evolution along the boundary of the first Brillouin zone \cite{XLQi2}. And it gives rise to the coefficient of the Hall conductor: $\sigma_{H}=C_1\frac{e^2}{h}$. Hence the ground state with non-zero $\chi$ processes the quantum anomalous Hall effect with the Hall coefficient $\pm \frac{e^2}{h}$. By contrast, the ground state with non-zero $\rho$ is a topologically trivial insulator.

\begin{figure}[t]
\centering
\includegraphics[width=3.1in]{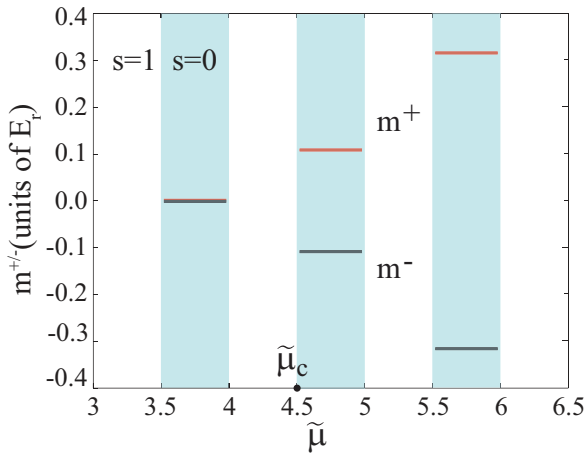}\hspace{0.5cm}
\caption{(Color online) The mass terms at the two Dirac points as a function of the filling of bosons $\widetilde{\mu}$. The composite fermion at the Dirac points gains the mass when the filling of bosons exceeds the critical value $\widetilde{\mu}_c$ about $4.5$. The parameters used here (in units of $E_r$, where $E_r=\hbar^2 k_r^2/2m$ is the recoil energy in experiment \cite{Bloch2}) are: $J_{F,1}=J_{B,1}=0.4$, $J_{F,2}=0$, $J_{B,2}=0.7$, $U=15$, and $\alpha=0.5$.
}
\label{fig:mass1}
\end{figure}

In the following, we consider the possibility of the realization of QAH in the Bose-Fermi mixture through the mean-field analysis. The values of the order parameters $\rho$, $\chi$, $\phi_A$ and $\phi_B$ are determined by the minimum of the free energy. At zero temperature, the free energy equals the ground state energy of the system $ \langle G | H_{eff}| G \rangle$. For example, we choose the interaction ratio as $\alpha=0.5$ and the chemical potential for bosons are tunable (as shown by the red line in Fig. \ref{phase}). The mass terms evaluated based on Eq. (\ref{eq:mass}) are shown in Fig. \ref{fig:mass1}. In the region with blue color, we give the values of the mass terms for the two Dirac points as functions of the filling of bosons $\widetilde{\mu}$. For $\widetilde{\mu}>\mu_c\approx 4.5$, the mass terms of the two Dirac points have non-zero values and process different signs. The Chern number is therefore equal to $1$ based on Eq. (\ref{Chern}). The mixture is broken to the QAH phase spontaneously. Accordingly, the spectrum for the Bose-Fermi mixture will open a gap at the Dirac point as shown by Fig. \ref{spectrum} (b), which is plotted for the filling of bosons as $\widetilde{\mu}=4.8$. The mass term increases as the filling of the bosons is increasing. The region with white color is not considered because that the composite particle in this region is formed with $s=1$. For reasons given in Sec. III, the temperature fluctuation is much larger than the bandwidth for this case, hence the derivations at zero temperature is not suitable for this region.

\begin{figure}[t]
\centering
\includegraphics[width=3.in]{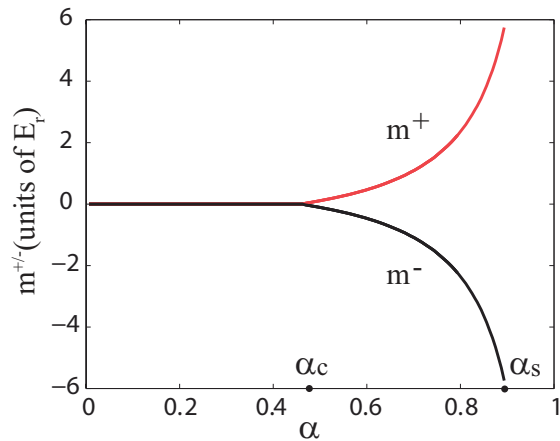}\hspace{0.5cm}
\caption{(Color online) The mass terms at the two Dirac points as a function of the interaction ratio $\alpha$. The composite fermion at the Dirac points gains a mass when $\alpha$ exceeds the critical value $\alpha_c$ about $0.48$, and this curve stops at $\alpha_s$. The filling of bosons is chosen as $\widetilde{\mu}=4.8$ and the other parameters used here are the same with Fig. \ref{fig:mass1}.
}
\label{fig:mass2}
\end{figure}

To see the effect of the interaction ratio $\alpha$, we plot the mass term as a function of $\alpha$ in Fig. \ref{fig:mass2}. We can see that the mass term obtains a non-zero value for $\alpha > \alpha_c \approx 0.48$. For bosonic fillings belonged to the same block of the $\alpha$-$\widetilde{\mu}$ plane, the critical point $\alpha_c$ is the same. This can be understood as the result of the fact that the phase boundary of the system is parallel to the $\widetilde{\mu}$ axes, which is shown explicitly in Fig. \ref{phase}. And the curves $m^{\pm}(\alpha)$ for different fillings belonged to the same block of the $\alpha$-$\widetilde{\mu}$ plane coincide for reasons that the effective model in each block is independent to $\widetilde{\mu}$ (see Eq. (\ref{eff_interaction})-(\ref{eff_hopping})). However, for $\alpha>\alpha_s$, where $\alpha_s$ is shown in Fig. \ref{phase}, the system will enter the region with $s=1$. The discussions above concentrate on the region with $s=0$ and are not applicable in this region with $s=1$. Hence the curve $m^{\pm}(\alpha)$ will stop at that point. There exist some regions of $\widetilde{\mu}$ where the value of $\alpha_s$ is lower than the critical value $\alpha_c$. These regions could be found in Fig. \ref{phase} by moving the green dashed line to left. Hence, for this case, the mass term is kept zero until the point $\alpha_s$ and the QAH effect would not appear.

To give a full description of the ground state property, we present a phase diagram about the system for $0<\alpha<1$ shown in Fig. \ref{phase}. The phase diagram for $-1 < \alpha < 0$ is the same with this figure for reasons that, the effective model given in Eq. (\ref{eff_interaction})-(\ref{eff_hopping}) for composite particle with $s=0$ does not depend on the sign of the interaction ratio $\alpha$. The QAH effect is formed in the orange region, where the composite particle is formed with $s=0$. The phase boundaries are lines parallel to the $\widetilde{\mu}$ axes. The purple region with $s=1$ is not considered. In this work, we considered the ultra-cold atomic system and the spin components, which can be simulated by the hyperfine states of ultra-cold atoms, are not included. When we extend the discussions to the case including the internal hyperfine states, the quantum spin Hall effect is also expected to be realized within this scheme.

Finally, we want to stress that the lowest energy scale for the observation of the QAH effect is determined by the bandwidth and energy gaps at the Dirac points and has the same order as the hopping parameter $J$. The bandwidth is determined by the effective hopping terms given in Eq. (\ref{eff_hopping}), which have the order $J$ for $s=0$ considered in this paper. As a result, the scale for the bandwidth could be estimated as $J$. The energy gaps driven by the effective neighbor site interaction are essential for the realization of QAH effect. As a result of second order of perturbative expansion, the effective neighbor site interaction has the order $J^2/V$. However, it could be dramatically amplified through both the interaction ratio $\alpha=U/V$ and the filling fraction of bosons $\widetilde{\mu}$, i.e., the factor $4 \alpha^2/(1-\alpha^2)$ and $n (n+1)$ in the right-hand side of Eq. (\ref{UV}). Through tuning the interaction ratio $\alpha$ near $\pm 1$ with the method of Feshbach resonance, or loading bosons to the optical lattice, the effective interaction could be increased quickly. Fig. \ref{spectrum}, \ref{fig:mass1}, and \ref{fig:mass2} plot the energy spectrum and the mass terms at the Dirac points, from which we could find that the energy gaps are comparable with the bandwidth of the energy spectrum. Driven by the interaction and filling effect, the scale for the energy gaps are boosted to become order J, namely, the hopping parameter. The corresponding experimental signatures will be discussed in the following section.

\begin{figure}[t]
\centering
\includegraphics[width=3.15in]{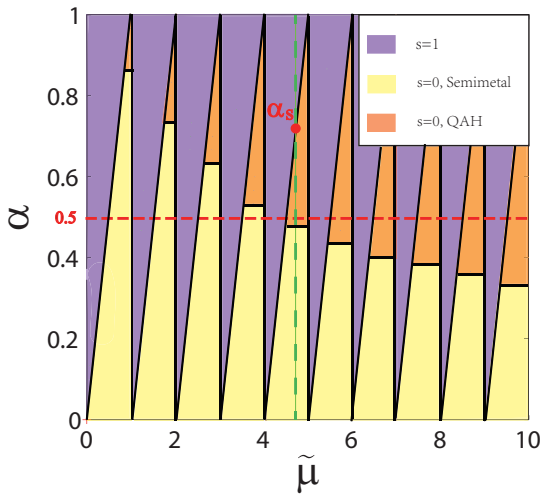}\hspace{0.5cm}
\caption{(Color online) The phase diagram with respect to the interaction ratio $\alpha$ and filling of bosons $\widetilde{\mu}$. The phase of the mixture for $s=0$ in the region $-1 < \alpha < 0$ is the same with the case for $0 < \alpha < 1$. The parameters used here are the same with Fig. \ref{fig:mass1}. On the blocks with orange color, the QAH state is formed, with the pase boundary parallel to the $\widetilde{\mu}$ axes. The red dashed line represents the case we considered in Fig. \ref{fig:mass1} with $\alpha=0.5$. The green line is the schematic of the path for the $\alpha$ axes in Fig. \ref{fig:mass2}. The red point denotes the place where the system enters the region with $s=1$ and the corresponding interaction ratio is denoted as $\alpha_s$.
}
\label{phase}
\end{figure}
\section{Experimental signatures and summaries}
In this paper, we have studied the possibility of the realization for QAH effect in Bose-Fermi mixture. To make contact with experiments, three practical considerations warrant mension: (i) In order to realize the independent trapping of the bosons and fermions, we could make use of the method of constructing the species-dependent optical lattice. The frequencies for the laser beams used to create the species-dependent optical lattice should be the same to ensure the coincidence of the two optical lattices. Therefore we could apply two sets of laser beams with the same frequencies but different polarizations to construct the optical lattices for the fermions and bosons independently. In recent years, the double species trapping of atomic gases \cite{Thalhammer,Modugno}and species-dependent optical lattice \cite{Cazalilla,WVLiu} have been realized in experiments. (ii) The validity of the effective model in Eq. (\ref{eff_interaction}) and (\ref{eff_hopping}) relies on the fact that $J \ll U,V$. The condition can be fulfilled for sufficiently strong optical lattice potentials, which has been accessed in the experiment for the observation of the Mott insulator \cite{Greiner}. (iii) The analysis for properties of the ground state of the Bose-Fermi mixture at $T=0$K is valid for $T$ much lower than the smallest energy scale in this system, i.e., the hopping parameters. This regime is accessible for sufficiently large interactions, which could be tuned by means of the Feshbach resonance \cite{Lewenstein,FeshbachLiRb}. In practice, the scale for the temperature fluctuation of the Bose-Fermi mixture ranges from nK to $\mu$K, while the corresponding energy scale of the $s$ wave scattering with the tuning of Feshbach resonance ranges from about $\mu$K to mK \cite{Bloch2,Feshbachcoldatom}, which covers about three orders of magnitude. Thus it is accessible, however still challenging, to tune the hopping strength $J$ by changing the potential depth of the optical lattice to fulfill the condition $k_B T \ll J \ll U,V$, where $k_B$ is the Boltzmann constant. Due to the stability of the topological properties of the QAH effect \cite{Kane2}, the first condition for $k_B T \ll J$ is not as strong as the second one $J \ll U,V$.

Up to now, the composite fermions proposed in Ref. \cite{Lewenstein} have been generated in experiment \cite{Sugawa}. The formation of the composite particle is a key demanding of our proposal, and this experimental realization of it provides the possibility for the realization of QAH in the Bose-Fermi mixture on the hexagonal lattice. In their experiment, the temperature of the cloud of $^{174}$Yb-$^{173}$Yb Bose-Fermi mixture ranges from $5$nK to $40$nK for repulsively interacting cases. And the energy scale for the background $s$-wave scattering of $^{174}$Yb-$^{173}$Yb atoms is about $k_B \times 168$nK. With the method of Feshbach resonance, the $s$-wave scattering strength could be further increased to fulfill the requirement of strong interaction. The detection of the QAH state in ultracold atomic systems has been presented extensively \cite{Zhusl}. The Bragg scattering \cite{BraggExp} and the standard density-profile technique based on the Streda formula \cite{Shao,Umucallar} are applicable in this experiment.

In conclusion, we studied the QAH effect in ultra-cold atomic system using a mixture of bosons and fermions loaded on the hexagonal optical lattice. The QAH effect is predicted to be formed when the filling of bosons is larger than the critical value $\widetilde{\mu}_c$. We have discussed the related experimental signatures for the realization of the QAH effect such as the construction of the species-dependent optical lattice, the Feshbach resonance in the Bose-Fermi mixture, and the temperature effects. Given the recent experimental developments on Bose-Fermi mixture and optical lattice, we expect that the proposal presented in this paper provides an alternative scheme to observe the QAH effect and test our understanding of its essence.

\begin{acknowledgments}
We acknowledge insightful comments by Jinwu Ye. This work was supported by the NKBRSFC under grants Nos. 2011CB921502, 2012CB821305, 2009CB930701, 2010CB922904, NSFC under grants Nos. 10934010, 11228409, 61227902 and NSFC-RGC under grants Nos. 11061160490 and 1386-N-HKU748/10.
\end{acknowledgments}

\end{document}